\documentstyle[prd,aps]{revtex}
\def\lb{\left(}
\def\rb{\right)}
\def\be {\begin{equation}}
\def\ee {\end{equation}  }
\def\beq{\begin{eqnarray}}
\def\eeq{\end{eqnarray}  }
\def\bi {\begin{itemize} }

\def\ei {\end{itemize}   }
\def\RE {I\kern-6pt R    }
\def\Z  {Z\kern-13pt Z   }
\def\be {\begin{equation}}
\def\ee {\end{equation}  }
\def\beq{\begin{eqnarray}}
\def\eeq{\end{eqnarray}  }

\def\eeq{\end{eqnarray}  }

\input epsf

\begin{document}
\draft

\twocolumn[\hsize\textwidth\columnwidth\hsize\csname
@twocolumnfalse\endcsname

\title{Critical Phenomena in Nonlinear Sigma Models}
\author{Steven L. Liebling and Eric W. Hirschmann}
\address{Theoretical and Computational Studies Group\\
     Southampton College, Long Island University,
     Southampton, NY 11968}
\author{James Isenberg}
\address{Department of Mathematics\\
         University of Oregon,
         Eugene, OR 97403}

\maketitle

\begin{abstract}
We consider solutions to the nonlinear sigma model (wave maps) with
target space $S^3$ and base space $3+1$ Minkowski space, and we find
critical behavior separating singular solutions from nonsingular
solutions. For families of solutions with localized spatial support a
self-similar solution is found at the boundary.  For other families, we
find that a static solution appears to sit at the boundary.
This behavior is compared to the
black hole critical phenomena found by Choptuik.
\end{abstract}


\vskip2pc]

\section{Introduction}
Nonlinear sigma models have been of considerable interest to both
physicists and mathematicians for a number of years. Physicists use them
to model symmetry breaking in the study of pions and other fundamental
particles, and also use them to model cosmological structure formation.
Mathematicians, who call them wave maps, use them as geometrically
motivated, nonlinear systems of hyperbolic partial differential equations
with which to study the formation and avoidance of singularities.

During the past ten years, mathematicians have proven first, that for
``small data", solutions of the Cauchy problem for wave maps avoid
singularities and exist for all time (``global existence")~\cite{small
data1,small data2}. They have also been able to show that, for
three or more spatial dimensions (in the base manifold), there are sets of
initial data which become singular in finite time~\cite{singular}. (In
one spatial dimension, this cannot happen~\cite{1dim1,1dim2}; it
is not yet clear whether singularities form in two spatial dimensions.)

These results together suggest that it could be interesting to consider
one-parameter families of initial data such that for small parameter
values no singularities occur, while for large values of the
parameter the fields become singular. Studying the evolution of such
families, one expects to see critical behavior of some sort occurring
near the transition values of the parameter. The recent work by Choptuik
~\cite{choptuik} and others in which experiments such as these have been
carried out with gravitational systems--collapse to a black hole for
large parameter values, and dispersal for small values of the
parameters--shows that very interesting phenomenology can be found at the
critical, transitional, values of the parameters.

Using primarily numerical methods, we carry out such studies for
spherically equivariant nonlinear sigma models with $S^3$ target
(corresponding to the symmetry breaking $SO(4)\rightarrow SO(3)$). We
find critical behavior which is similar in some ways to that seen by
Choptuik and collaborators, but very different in other ways.

We first focus on sets of initial data with localized support and finite
energy. For families of such solutions, the small data global existence
results hold for small values of the parameters, and the presence of
critical behavior is unambiguous. We find for these families a
unique, continuously self-similar solution at the threshold. This critical
solution is an intermediate attractor so the critical behavior is
``type II", like that seen in critical collapse to a black hole.

Motivated by the Turok-Spergel solution  ~\cite{turok}, which is the
only known explicit wave map solution which evolves from regular initial
data to a singularity, we also examine sets of initial data which do not
have localized support and have infinite energy. Although the small data
theorem does not apply to solutions generated by such data, the
``texture" studies of wave maps suggest
that both nonsingular and singular solutions should occur~\cite{peri,sornborger2}.
Our numerical
studies support this contention, and we have found that the transition is
marked in some cases by the self-similar solution noted above, but in
others by static solutions. While the static solutions we see at the
transition are not intermediate attractors, and therefore are not
critical solutions in the usual sense, our studies indicate interesting
behavior which deserves further exploration.

We describe in more detail what we have learned about critical
and threshold behavior for $S^3$ wave maps in sections III and IV. Before
doing this, we  briefly review in section II what wave maps are, the
equations for spherically equivariant wave maps, and some of
the families of initial data we use to probe the critical boundary. We
describe the results of these numerical probes in section III for the
families of data with localized support,and in section IV
for the other families. We also note in III and IV  some of the
properties of the solutions found on this boundary. We make a few
concluding remarks in section V.

\section{Spherically Equivariant Wave Maps}

A nonlinear sigma model, or wave map, is defined to be a map  $\phi^a$
from a (Lorentz signature) spacetime (the ``base") into a Riemannian
geometry (the ``target"), with the map satisfying the differential equation

\be
    \partial^\mu \partial_\mu \phi^A
  + \Gamma^A_{BC} \partial_\mu \phi^B \partial^\mu \phi^C = 0
\ee
where $\Gamma^A_{BC}$ represents the Christoffel symbols corresponding
to the metric on the target space.

In this work, we fix the base to be 3+1 Minkowski spacetime, and we
fix the target to be $S^3$. Furthermore, we make the spherical equivariance
ansatz, which may be expressed in ``hedgehog" coordinate form
(for $S^3 \subset R^4$) as follows
\be
\phi^a = \left( \begin{array}{c}
                \sin \chi(r,t) \sin \theta \sin m \varphi \\
                \sin \chi(r,t) \sin \theta \cos m \varphi \\
                \sin \chi(r,t) \cos \theta \\
                \cos \chi(r,t)
                \end{array} \right),
\label{eq:phi}
\ee
with $m$ a positive integer.

The only free function in~(\ref{eq:phi}) is the spherically
symmetric function $\chi(r,t)$. It satisfies the nonlinear wave equation
\be
  \ddot \chi
- \frac{1}{r^2} \left( r^2 \chi' \right)'
= - m(m+1) \frac{\sin \left(2\chi\right)}{2r^2},
\label{eq:eom}
\ee
where prime denotes $\partial/\partial r$ and an overdot
denotes $\partial / \partial t$.
We enforce the regularity condition $\chi(0,t)=0$ at the origin,
and apply a standard out-going radiation boundary condition
at large radius.
The radial energy density corresponding to this
system is
\be
\rho(r,t) = \frac{r^2}{2} \left[
                                   \dot \chi^2
                                 + \lb\chi'\rb^2
                                 + \frac{m(m+1)}{r^2}\sin^2 \chi \right],
\ee
with the corresponding  energy function
\be
E(t) = \int_r \rho(r,t)~dr.
\ee

One of the features of this spherical equivariance ansatz is the
possibility of nontrivial ``texture charge" or ``degree". The degree of a
particular wave map $\phi(r,\theta,\varphi,t_0)$ at a fixed time $t_0$
corresponds to the multiplicity of the covering of the target sphere
$S^3$ (ie, the order of the third homotopy group). In terms of the
hedgehog form~(\ref{eq:eom}), the degree depends on $m$, on the range of
$\chi(r,t_0)$, and on certain continuity conditions at the poles of $S^3$.
We note that the degree is zero so long as the range of $\chi(r,t_0)$ is
less than $\pi$; if the range of $\chi(r,t_0)$ is greater than $\pi$,
the degree may or may not be nonzero. The degree does not change during
a smooth evolution.

If the degree of a wave map is nonzero, the energy cannot be
arbitrarily small. Hence, small data arguments for global existence
cannot be used. Indeed, numerical evidence (ours and that of others)
suggests that degree nonzero wave maps are inevitably singular. While
this has not been proven, it leads us, in studying criticality, to
focus on zero degree initial data.

To fully specify initial data, we must specify both $\chi(r,0)$
and its time derivative at the initial time, $\dot \chi(r,0)$.
We then evolve this initial data with a first order formulation
in which we take our fundamental fields to be $\chi(r,t)$ and
$\Pi(r,t) \equiv \dot \chi(r,t)$. As a matter of convenience,
we generally take as initial data $\Pi(r,0)=\chi'(r,0)$ such that
the field $\chi(r,0)$ represents an approximately in-going pulse.
This choice has no affect on the critical behavior but helps to
mitigate reflection from the outer boundary.
Our method makes use of
an iterative, second order accurate, Crank-Nicholson finite difference
scheme which we have incorporated into the adaptive framework developed
by Choptuik~\cite{choptuik}. We have tested this code and shown it to
converge quadratically,
to conserve energy, and to be stable.

The first two families of initial data we have used to probe criticality
have been chosen to have localized support and finite energy.  The
parameters in these families can be chosen so that the energy is very
small, in which case the small data global existence results guarantee
that no singularity will develop. For other parameter values, the energy
is large, and the development of singularities is expected. Note that in
each case, there is an amplitude $A$ which we use to scale the data from
nonsingular to singular solutions, and in addition there are two other
parameters
$R_0$ and
$\delta$ which we can use to change some of the qualitative features of
the family:\\
{\em Gaussian Pulse Data}
\beq
\chi(r,0) & = & A e^{-\left(r-R_0\right)^2/\delta^2}  \cr
\Pi(r,0)  & = & \chi'(r,0),
\label{eq:zero_id}
\eeq
{\em Logarithmic Data}
\beq
\chi(r,0) & = & A \frac{\ln \left( r + R_0 \right)}{r + \delta} \cr
\Pi(r,0)  & = & \chi'(r,0)
\label{eq:logr}
\eeq

We also examine two other families of data. One of them is special in
that it includes the initial data (for
$\epsilon=1$) which generates the Turok-Spergel solution~\cite{turok}.
This is the explicit self-similar solution which is known to evolve into a
singularity in finite time. Note that the Turok-Spergel solution has
nonzero degree; all others in this family (with $\epsilon < 2$) have zero
degree. Note also that the energy for all of the data in this family is
infinite. Therefore, we cannot use the small data global existence
theorem to guarantee that solutions generated by data with $\epsilon$
small will be nonsingular. This does, however, appear to be the case
(see section IV). The same is true for our fourth family of data;
while there is no theoretical guarantee that both nonsingular and
singular solutions are generated by data in this family, our numerical
evidence supports the contention that both do occur, and so transition
behavior can be studied.

 {\em Generalized Turok-Spergel
Data}~\cite{turok}
\beq
\chi(r,0) & = & 2 \epsilon \tan^{-1} \left( \frac{r}{\Delta} \right) \cr
\Pi(r,0)  & = & \frac{2\epsilon r}{\Delta^2 + r^2}.
\label{eq:ts_id}
\eeq

{\em Tanh Data}
\beq
\chi(r,0) & = & A \left[ \frac{1}{2} \tanh\left( \frac{r-R_0}{\delta} \right)
                        +\frac{1}{2} \right] \cr
\Pi(r,0)  & = & \chi'(r,0).
\label{eq:tanh}
\eeq

For each of the families of data listed above, our numerical studies
proceed as follows: We fix a specific family by fixing a choice of $R_0$
and $\delta$ (or a choice of $\Delta$ ) in one of the family classes
listed above. With that fixed family, we run through a number of choices
of $A$ (or $\epsilon$), from very small to large, and we evolve the
solution for each choice.

In the evolved solutions, we carefully monitor the
behavior of the energy density function $\rho(r,t)$ as well as that of
$\chi(r,t)$; and we use these behaviors to determine which solutions become
singular and which do not.
Singularity formation is indicated by the unbounded growth of the
derivatives $\chi'$ and $\dot \chi$ (and hence $\rho/r^2$).
We find in each case that there is a critical value of $A$ (or
$\epsilon$) which divides the initial data that evolve into a singularity
from those
which do not. We study very carefully the solutions at or near this
critical value.

We note that while numerical results
never prove that a solution is singular or not, in these studies the
singular behavior appears dramatically as much of the energy density
concentrates and grows without apparent bound at the origin. Note that for all
of the solutions, the energy density initially flows towards the origin.
In the nonsingular cases, the energy density grows at the origin, and
then disperses; while in the singular cases, it continues to grow.

In the course of our studies, we have  noticed another useful signal of
impending singular evolution: In all cases, whenever the range of
$\chi(r,t)$ exceeds $\pi$ at a given time, a singularity occurs to the
future. Whether or not one can indeed prove such a result, it is useful
in sorting the evolutions.

\section{Self-Similar Solutions at Criticality}

\begin{figure}
\epsfxsize=8cm
\centerline{\epsffile{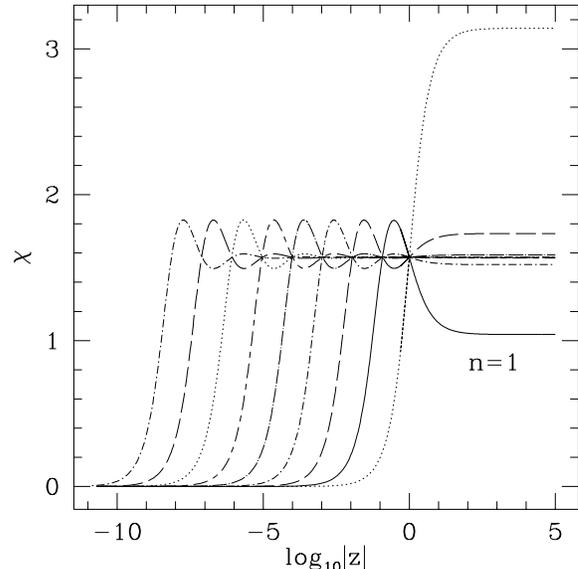}}
\caption{These are the first nine members of the family of self-similar
         solutions found by Aminneborg and Bergstrom~[8]
         as well as Bizon~[9].
         The $n=0$ solution is the original
         Turok-Spergel solution while the $n=1$ solution (the solid line)
         is the critical solution which serves as an intermediate attractor
         for the collapse of certain families of initial data.  The label $n$
         labels the number of times $\chi$ crosses the line $\pi/2$ on the
         interval $(0,1)$.
         }
\label{fig:AB_solns}
\end{figure}

If we consider Gaussian Pulse data for various fixed values of $R_0$ and
$\delta$, we find that as $A$ approaches its critical value $A^{*}$, the
corresponding solution approaches a particular self-similar solution.
This critical solution is {\it not} the solution found by Turok and Spergel;
rather it appears to be one of the sequence of self-similar solutions
discovered by Aminneborg and Bergstrom~\cite{aminneborg}, and subsequently
Bizon~\cite{bizon}.  These regular,
self-similar solutions obey~Eq.(\ref{eq:eom})
together with the scaling assumption that $\chi(r,t) = \chi(-r/t)$.
The resulting equation is
\be
z^2 (z^2-1) \chi_{,zz} + 2 z (z^2-1) \chi_{,z} + \sin(2\chi) = 0
\ee
where the differentiation is with respect to $z\equiv -r/t$.
This equation admits a countably infinite number of solutions, labeled by $n$,
the number of times the solution crosses $\pi/2$ between $z=0$ and $z=1$.
The Turok-Spergel solution is the $n=0$ solution. For all $n$ except $n=0$,
these solutions have zero texture charge.
Figure~\ref{fig:AB_solns}
plots the first several solutions in this family.

Within the range of values of
$r$ in which the near-critical solutions approach self-similarity, it
is not easy to distinguish the various members of the AB sequence of
solutions. In order to determine which of these solutions, which we will call
$AB_n$, does
occur on the boundary between singular and nonsingular
solutions, we have examined the behavior of solutions near to several of the
$AB_n$ solutions. Specifically, on evolving members of this family, we
choose a time $t_0$ and add a parametrized set of small amplitude
Gaussian pulses to the exactly self-similar solution as initial data
at $t_0$ for $AB_n$. Only for
$AB_{1}$ do we find that for negative amplitude pulses, the
solution is nonsingular while for positive amplitude pulses, the
solution is singular. This is particularly convincing evidence
that $AB_{1}$ is the critical
solution, and the others are not.

In addition to the nonlinear evolution of these self-similar solutions, we
obtain further confirmation that $AB_{1}$ is the critical solution
by carrying out a linear perturbation analysis for it, as well as
for some of the other $AB_n$ solutions.

Our linear perturbation analysis around this family of self-similar solutions
is standard.  In coordinates adapted to the self-similarity
($z\equiv -r/t$ and
$\tau\equiv \ln|-t|$), the perturbed solution to leading order will be
\be
\chi(r,t) = \chi_{0}(z)
      + \delta\cdot\int e^{\lambda\tau} \hat{\chi}_{1}(z;\lambda) d\lambda
\ee
where $\chi_{0}(z)$ refers to any member of the $AB_n$ family and
$\hat{\chi}_{1}$
is an eigenmode of the perturbation expansion associated
with the eigenvalue $\lambda$.
With this expansion, the eigenmodes obey the linear equation
\beq
    z^2(z^2-1)\hat{\chi}_{1,zz}
  + 2z\left(z^2-1-\lambda z^2\right)\hat{\chi}_{1,z} \cr
  + \left(2\cos(2\chi_0) + \lambda^2 z^2 - \lambda z^2\right)\hat{\chi}_{1}
& = & 0.
\eeq
In general, $\lambda$ can be complex, but in this case it will suffice to
consider $\lambda$ real.  As $t\rightarrow0$,
$\tau\rightarrow-\infty$, thus if $\lambda>0$, the corresponding perturbations
will decay.  However, if $\lambda<0$, the perturbations will grow and render
the original self-similar solution unstable.

In order to solve the above equation it is sufficient to demand regularity
at $z=0$ and $z=1$.  On performing the integration, we find that there is
a single gauge mode at $\lambda=-1$ for all members of the $AB_n$ family.  This
gauge mode arises because of the freedom we have in choosing the zero of time:
$t\rightarrow t+c$.  In addition to this gauge mode we confirm that the
Turok-Spergel solution (the $n=0$ member of this family) has no
unstable modes, the $n=1$ member of this family has a single
unstable mode, and that for all the exactly self-similar solutions we have
considered with $n>1$,
there always exists more than a single unstable mode~\cite{footnote}.

Thus this serves as further evidence that $AB_{1}$ is the critical
solution.  In the sense of dynamical systems, that this exactly self-similar
solution has a single unstable mode indicates that it is an intermediate
attractor on the boundary between the basin of attraction for singular
solutions and the basin of attraction for nonsingular solutions. When
such an attractor exists for critical behavior, one is said to have a
``type II transition."

For the case of this intermediate attractor, the $AB_{1}$ solution, the
eigenvalue for the single unstable mode is found to be $\lambda \approx -6.33$.

\begin{figure}
\epsfxsize=8cm
\centerline{\epsffile{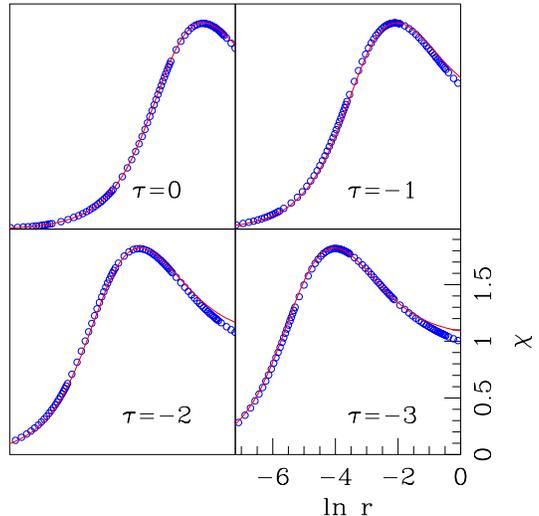}}
\caption{Demonstration of the self-similarity of the critical solution
         using initial data of the form~Eq.(\ref{eq:logr}). Letting
         $\tau \equiv \ln |T^*-T|$ where $T$ is the time of collapse,
         the four frames are equally spaced in ``log time'' progressing
         towards collapse ($\tau\rightarrow -\infty$).
         A near-critical solution for $\chi(r,t)$ is shown (circles) for
         $R_0=1$ and $\delta=1$
         versus $\ln r$. The $n=1$ self-similar solution is shown (solid)
         with the freedom to set the collapse time used to make the
         two solutions coincide in the first frame only.
         That the solutions coincide at
         the other times demonstrates that
         the critical solution is self-similar and approaches
         the $n=1$ solution.
         }
\label{fig:ss}
\end{figure}

In general,
we find the same critical behavior occurring at the transition for all
families of Gaussian Pulse data that we have evolved. In addition, for a
number of families of Logarithmic data and even some families of the
nonlocaly supported Tanh data, we find
$AB_{1}$ occurring at the transition as well. ( Figure~\ref{fig:ss}
indicates the closeness of the
evolution of a near critical solution for Logarithmic data and the
evolution of $AB_{1}$.) This suggests that
$AB_{1}$ is, at least in a local sense, ``universal". Universality is a
familiar occurrence in nonlinear dynamics. For example, for a damped
pendulum, for all initial data except that corresponding to the
stationary straight up position, the pendulum eventually ends up in the
stationary straight down state. This down state is a universal attractor for the
whole system.

A particularly pertinent example of similar behavior has been found in the
study of black hole collapse critical behavior (for a review
see ~\cite{review1,review2}). This work has demonstrated
that gravitational collapse exhibits critical solutions at the threshold
of black hole formation.  There, the exactly critical solution within a
specific model exhibits
universality as well as self-similarity (which, depending on the
model investigated, can be discrete or, as here, continuous).  The
gravitational critical solutions are also intermediate attractors,
like the $AB_{1}$ solution, in that they have a single unstable mode and
sit on the boundary between the dispersal of the collapsing matter and the
formation of a black hole ({\it i.e.} singularity).  Presumably,
if we were to couple this nonlinear sigma
model to gravity and evolve similar initial
data, we would get black hole formation.  But what is especially significant
here is that even without gravity, we get singularity formation together
with the universality and self-similarity seen in the gravitational
context.

\section{The Role of Static Solutions}

Consider now evolving the Generalized Turok-Spergel data for various values of
$\epsilon$.
By fixing a value of $\Delta$ and considering
solutions parametrized by $\epsilon$
we might expect to again get critical behavior as before.
Though we do observe some sort of threshold behavior, the $AB_{1}$
self-similar solution does {\em not} occur at the transition.
Instead, our numerical evolutions suggest that static solutions
play a role in the threshold behavior.

The possibility that static solutions occur
at the transition between singular and
nonsingular data has led us to consider whether
the static solutions are critical in this sense. To investigate this
possibility, let us first consider the stability properties of the
strictly static solutions.

Static solutions are studied in~\cite{lichtensteiger} and here we
consider only those for which $\chi(0)=0$. We could parametrize
this family by $a \equiv \chi'(0)$,
however, Lichtensteiger and Durrer observe that the static
solutions are all related by a simple radial rescaling so we need consider
only $a=1$.
We consider initial data of the form
\beq
\chi(r,t) & = & \chi_s(r) + A e^{-(r-R_0)^2/\delta^2} \cr
\Pi(r,0)  & = & \left[ -\frac{r-R_0}{\delta^2} \right]
                 A e^{-(r-R_0)^2/\delta^2}
\eeq
which we proceed to evolve. The above initial data represents the
static solution, $\chi_s(r)$, perturbed by an in-going Gaussian pulse.
As with our non-linear perturbation of the self-similar solution, our
expectation is that threshold behavior would be demonstrated
if for $\chi_s(r)$ the solution
becomes singular for positive amplitude perturbations ($A>0$) but remains
nonsingular for perturbations with $A<0$.
A similar test is used
in~\cite{vanputten,liebling}
to determine whether static solutions sit on the threshold of
black hole formation.

\begin{figure}
\epsfxsize=8cm
\centerline{\epsffile{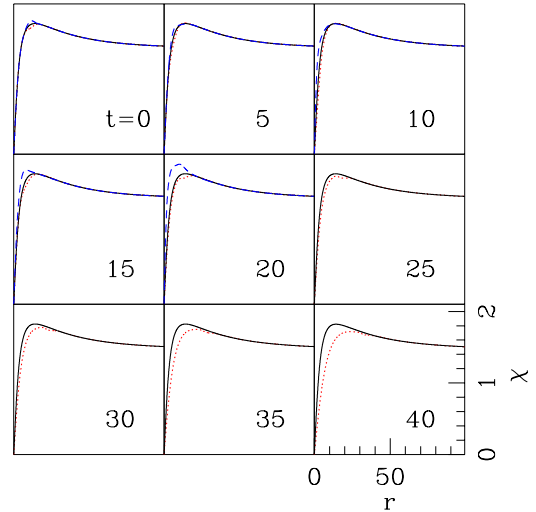}}
\caption{Demonstration of the instability of the static solution.
         The static solution (solid) is perturbed with a positive
         amplitude Gaussian pulse (dashed) and negative amplitude (dotted).
         The positive perturbation collapses while the negative one
         disperses suggesting that the static solution sits on threshold.
         }
\label{fig:static_unst}
\end{figure}

An example of this experiment is shown in Fig.~\ref{fig:static_unst}.
The figure demonstrates that
nonlinear perturbations of opposite sign send the static solution either
to collapse or dispersal depending on the sign of the perturbation. This
suggests
that the static solution $\chi_s(r)$ does indeed sit on threshold.

However, if the static solution sits on threshold, one would expect that
it has a single
unstable mode. If so,
then it should be an intermediate attractor within some
basin of attraction. If it has more than one unstable mode,
then we would not expect to find it via a one parameter tuning.

The mode structure for the static solution is determined by doing a
linear perturbation study. Writing
\be
\chi(r,t) = \chi_s(r)
        + \delta \cdot \int e^{-i\omega t} \tilde{\chi}_1(r;\omega) d\omega,
\label{eq:pert}
\ee
with $\chi_s(r)$ denoting the static solution, with
$\tilde{\chi}_1(r;\omega)$ denoting the perturbation, and with
$\omega$ being the eigenvalue associated with the perturbative mode
$\tilde{\chi}_1$, we determine  (after substituting~(\ref{eq:pert})
into~(\ref{eq:eom})), that the perturbation modes obey
\be
\tilde{\chi}_1'' = \frac{2\tilde{\chi}_1 \cos \left( 2 \chi_a \right)}{r^2}
          - \omega^2 \tilde{\chi}_1 - \frac{2}{r} \tilde{\chi}_1'.
\ee
with the regularity conditions
\be
\tilde{\chi}_1(0) = 0 ~~~~ \tilde{\chi}_1'(0) = {\rm free}.
\ee
Unstable perturbation modes are signaled by $\omega^2<0$.
We find solutions numerically, using  a standard shooting technique
with the   regularity condition at infinity being
$\tilde{\chi}_1'(r\rightarrow\infty)=0$. Due to the linearity of the
problem, we let $\tilde{\chi}_1'(0)=1$ and adjust $\omega^2$ until our
regularity conditions are met.
We find a  number of unstable ($\omega^2<0$) modes; it follows
that  the static solution does not represent an
intermediate attractor.

Since the static solution is clearly not an intermediate attractor,
 we might not expect it to be found by tuning
the Generalized Turok-Spergel initial data. For this reason, we
do not view the static solution as a critical solution in the usual sense.
However, it does seem to occur at the threshold, both for the Generalized
T-S initial data and for a number of families of Tanh data.

Here, we might comment on some of the difficulties associated with the
numerical study of solutions generated from data with infinite support:

As stated previously, fixing $\Delta$ and picking a large $\epsilon$
for the TS data, the evolution clearly demonstrates singular collapse
(for $\epsilon=1$ collapse is known).
With a small value of $\epsilon$, one might expect to observe
dispersal. That is, one might expect to observe some
energy density initially moving towards the origin, turning around, and then
traveling outwards forever. The problem here is that numerically we can
neither evolve forever nor evolve over an infinite domain.
Our evolutions are limited in domain because of finite computer resources
and limited in time by the adulteration of boundary effects exacerbated
by the infinite nature of the initial data.

The problem of determining dispersal
is less crucial  for the case of Gaussian initial data and other families
with localized support because we can rely on the small data global
existence theorems to guarantee that the evolution is nonsingular.
Here though, those theorems are not applicable because the initial
data has infinite energy.

As corroborative support for our view that we are seeing nonsingular
solutions, we note the work on textures in which  scaling
arguments are used to show that, at least for a particular class of
initial data of infinite support, wave map evolutions that do not
collapse can occur~\cite{peri}. In fact,  a number of these papers
discuss the critical winding number of such textures (of infinite
support) which separates dispersal from collapse (see, for
example,~\cite{sornborger2}).

Our evolutions for small $\epsilon$ show what appears to be
dispersal, and those for large $\epsilon$ show apparent collapse. Tuning,
however, is very difficult, since (as seen in ~\cite{sornborger2}),
we observe solutions which at first appear to be dispersing but ``turn
around" and then ultimately collapse.  This turn-around can occur very
slowly. Hence, finding the transition is very hard.

Our evolutions therefore suggest three regimes in $\epsilon$
for the TS initial data, as well as for
certain families of the Tanh data~\cite{footnote3}.
For large $\epsilon$, the evolutions quickly
collapse. For small $\epsilon$, the evolutions suggest that the
solutions do not collapse but instead disperse. For moderate $\epsilon$,
solutions appear to be dispersing but then turn around and collapse.

\begin{figure}
\epsfxsize=8cm
\centerline{\epsffile{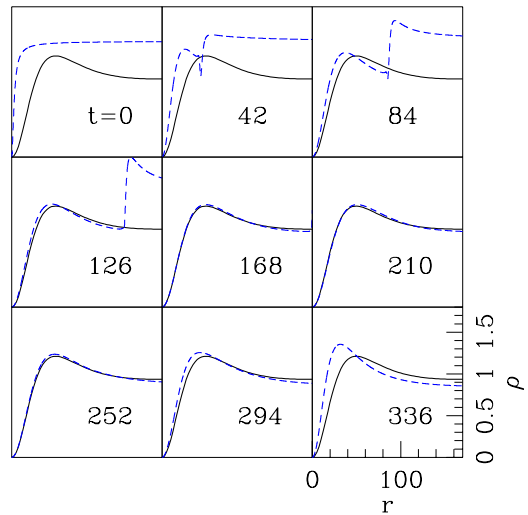}}
\caption{Apparent approach of the tuned Turok-Spergel
         initial data to the static solution. Shown (dashed) is the
         evolution of the energy density $\rho(r,t)$ for the
         Turok-Spergel initial data, Eq.(\ref{eq:ts_id}),
         with $\epsilon=0.302$.  The energy quickly begins to
         move outward to large
         $r$. However, by $t\approx 126$, the evolution
         has shed a large component of its energy density leaving
         behind an approximately static solution. Shown also
         is the energy density for the $a=0.12$ static
         solution (solid),
         chosen for the best correspondence to the static part
         of the evolution.
         }
\label{fig:static_tuned}
\end{figure}

Given the resulting difficulty in finding a critical $\epsilon$, one is
led to ask in what sense the static solution exhibits threshold behavior.
It seems to arise for the intermediate range of $\epsilon$
as an evolving solution ``turns around" from its
initially outgoing, dispersive behavior and begins its collapse to a
singularity.  As this ``turn around" point is approached, the field
profiles approach that of the static solution and remain there for
a certain amount of time~\cite{footnote4}.
An example of this is shown in Fig.~\ref{fig:static_tuned}.
Although this behavior is certainly reminiscent of observed critical
behavior, since the static solution has multiple unstable modes,
it is not an intermediate attractor, and so not a critical solution in
the accepted sense.  However, it appears that this static solution does
arise in some sense, and does play some role
in wave map threshold behavior.

The above discussion simply describes what our numerical evolutions
suggest, but is clearly not definitive.
Nonetheless, we conjecture that while for small $\epsilon$ data
the solutions do disperse, for somewhat larger $\epsilon$ data, the
solution will appear to be dispersing, but then will approach the static
solution
$\chi_s(r)$ (in general, for some $\epsilon$-dependent $a$), and will finally collapse.
Further, we conjecture that as one decreases $\epsilon$, one will observe
the solution turning around at later and later times (and larger and larger
$r$) until for some non-zero value, $\epsilon^{*}$, the solution
``turns around'' at infinite time and radius.  Any further reduction of
$\epsilon$ below $\epsilon^{*}$ results in dispersal.

\section{Conclusion}

Our work shows that nonlinear sigma models, or wave maps, from $3+1$
Minkowski spacetime into $S^3$ exhibit critical behavior which is
similar to that seen in the study of black
hole collapse for Einstein's equations with various source fields. We
find that the boundary between sets of data evolving into nonsingular
solutions and sets of data evolving into singular solutions includes a
self-similar solution. The static solutions are found to play a role as
well. The self-similar solution is an intermediate attractor, while the
static solutions are not.

While this work is a first step toward understanding critical behavior
in wave maps, it leaves a number of questions unanswered:

1) Does the critical boundary for spherically symmetric wave maps from 3+1
Minkowski spacetime into $S^3$ include other solutions besides those we
have seen?

2) How do the solutions on this boundary fit together?

3) What happens if one removes the spherical equivariance condition?

4) What happens for target spaces other than $S^3$?

5) What happens for base spaces other than 3+1 Minkowski spacetime?

A base space of particular interest is 2+1 Minkowski spacetime. For 2+1
wave maps, it is not yet known whether in fact there any singular
solutions which evolve from regular initial data (2+1 is the ``critical
dimension" for the  wave map system of partial differential equations,
just as 4+1 is the critical dimension for Yang-Mills). If such solutions
exist, there would likely be critical behavior. However, one expects the
nature of the critical boundary between singular and nonsingular
solutions to be very different in this case. This issue is currently
under study.

\section*{Acknowledgments}

While this work was in preparation, a preprint by
Bizon~\cite{bizon} appeared which referred to some recent results of his and
collaborators's and which has some
overlap with the work described here.
We are grateful for the hospitality of the ITP  (supported in
part by the National Science Foundation under Grant No. PHY94-07194)
at the University of California, Santa Barbara
at whose conference,
{\em Classical and Quantum Physics of Strong Gravitational Fields},
this work began.
Partial support for this work has come from NSF Grant
PHY-9800732 at the University of Oregon. SLL and EWH are also appreciative
of the financial support of Southampton College.



\begin{references}
\bibitem{footnote}
   We conjecture, as do
   Bizon and his collaborators~\cite{bizon},
   that the $n^{{\rm th}}$ self-similar solution
   will have exactly $n$ unstable modes. However, we are unaware of any
   proof of this and though our integrations provide evidence to this end,
   they do not, unfortunately, constitute proof for all $n$.
\bibitem{footnote2}
   Because the static solution has many unstable
    modes, this test is only suggestive, not conclusive.
\bibitem{footnote3}
   We do not expect, nor do we observe, initial data families of compact
   support finding anything other than the $AB_1$ self-similar solution
   on searching.
\bibitem{footnote4}
   This time during which the evolution resembles the static solution
   does not appear related to tuning $\epsilon$. Because the static
   solution is not an intermediate attractor, more precise tuning of
   epsilon does not result in a longer time.
\end{references}
\end{document}